\documentclass{emulateapj}

\def\icarus{Icarus}
\usepackage{natbib}
\slugcomment{ApJ Letters, in press}

\begin{document}

\title{A hypothesis for the color diversity of the Kuiper belt}
\author{M.E. Brown}
\affil{Division of Geological and Planetary Sciences, California Institute
of Technology, Pasadena, CA 91125}
\email{mbrown@caltech.edu}
\author{E.L. Schaller}
\affil{NASA Dryden Aircraft Operations Facility, Palmdale, CA 93550
and
National Suborbital Education and Research Center, University of North
Dakota, Grand Forks, ND, 85202}
\author{W.C. Fraser}
\affil{Division of Geological and Planetary Sciences, California Institute
of Technology, Pasadena, CA 91125}

\begin{abstract}
We propose a chemical and dynamical
process to explain the surface colors of the Kuiper belt.
In our hypothesis, the initial bulk compositions of the bodies themselves
can be quite diverse -- as is seen in comets -- but the early surface
compositions are set by volatile evaporation after the objects 
are formed. Strong gradients in surface composition, coupled with
UV and particle irradiation, lead to the surface colors that are seen today.
The objects formed in the inner part of the primordial belt retain only
H$_2$O and CO$_2$ as the major ice species on their surfaces. 
Irradiation of these species plausibly results in the dark neutrally
colored centaurs and KBOs. Object formed further in the disk retain
CH$_3$OH, which has been shown to lead to brighter redder surfaces after
irradiation, as seen in the brighter redder centaurs and KBOs. Objects
formed at the current location of the cold classical Kuiper belt uniquely
retain NH$_3$, which has been shown to affect irradiation chemistry and
could plausibly lead to the unique colors of these objects. We 
propose observational and experimental tests of this hypothesis.
\end{abstract}

\keywords{solar system: Kuiper belt --- solar system: formation --- astrochemistry }

\section{Introduction}
Nearly twenty years after the realization that the region beyond Neptune is
a depository for vast numbers of objects, 
one of the first discoveries about the physical properties of these 
Kuiper belt objects (KBOs) -- that they span a wider range of colors than
most other solar system populations -- remains unexplained \citep{1998AJ....115.1667J, 1996AJ....112.2310L, 2008ssbn.book...91D}.
The surfaces of KBOs range from those which are neutrally reflecting -- and 
thus appear to have essentially solar colors -- to some of the reddest objects
known in the solar system. The full range of colors is mixed at
what appears to be nearly 
random throughout the outer solar system \citep{morbandbrown}.
Early hypotheses of randomized collisional excavation
 \citep{1996AJ....112.2310L} or velocity dependent impact resurfacing 
\citep{2002AJ....124.2297S} have proven incapable of 
reproducing the features of the observations \citep{2008ssbn.book...91D}, 
yet no alternatives have been proposed.

We propose a chemically and dynamically plausible hypothesis where 
the surface compositions and thus
colors of KBOs and their progeny are set by formation-location
dependent volatile loss in the early solar system. In the next section we
present the observational constraints on KBO colors. Next
we detail the volatile-loss gradient hypothesis , and finally, 
we present predictions for this hypothesis.

\section{Observational constraints}

While the causes of color variation among KBOs remain unclear, 
key observational constraints provide important clues
for understanding the surfaces of these objects.

 {\it Cold classical KBOS.}
While most of the Kuiper belt appears to be composed of essentially the
same mixture of gray to red objects \citep{morbandbrown,2008ssbn.book...91D}
one dynamical region stands out for its
homogeneous composition. The cold classical Kuiper belt was first identified
as a dynamically unique region of the Kuiper belt -- a difficult-to-explain 
overabundance of low inclination, dynamically cold objects beyond about
41 AU \citep{2001AJ....121.2804B}. 
Subsequent observations revealed that these objects
shared a common red coloring \citep{2002ApJ...566L.125T}. 
These cold classical KBOs are
now also know to be unique in their lack of large bodies 
\citep{2001AJ....121.1730L}, their
higher abundance of satellites \citep{2008Icar..194..758N}, 
and their different size distribution \citep{2010Icar..210..944F}.
Preliminary results also suggest that the cold classical KBOs have higher albedos than
those of the remaining population
\citep{2009Icar..201..284B}. 
All of these properties appear to signify a population with a 
different -- and perhaps unique -- formation location or history. 
Explaining the uniformly red color of these objects is critical to
any understanding of Kuiper belt surface colors.

 {\it Centaurs.} Centaurs -- former KBOs currently
 occupying short-lived giant planet-crossing orbits -- provide
a second important constraint on KBO colors. 
Centaurs are derived from the Kuiper belt and
should start with the surface chemistry range inherent in that
population.

While the range of centaur colors generally covers the full range of colors
covered by the Kuiper belt, the centaurs appear deficient in colors in
the middle part of the range, giving the distribution of centaur colors
a bimodal appearance \citep{2008ssbn.book..105T}. 
This bimodality has been hypothesized to
be due to a surface modification as centaurs are heated as they approach
closer to perihelion. None of the specific models proposed, however,
is able to fit the detailed spectral, albedo, and color observations 
\citep{2009AGUFM.P21B1213S}.

{\it Local conditions.}
The colors of scattered, resonant, and hot classical Kuiper belt objects
(that is, everything except cold classical Kuiper belt objects and centaurs)
are uncorrelated with any current dynamical parameter. This lack of
correlation, particularly with semi-major axis or perihelion/aphelion 
distance argues strongly that local heating, UV irradiation, 
and solar wind and cosmic ray
bombardment \citep{2003EM&P...92..261C} are not responsible for
the varying colors of the Kuiper belt. Local conditions appear to
have no primary influence on the colors of KBOs.

 {\it Binary KBOS.} Careful measurement of the colors of the separate 
components of binary KBOs has shown a tight correlation over the full
range of Kuiper belt colors \citep{2009Icar..200..292B}. 
The colors of two KBOs in orbit around
each other are almost always nearly identical.
This fact immediately rules out stochastic processes such as
collisions for the causes of these Kuiper belt colors. Indeed, 
given the lack of correlation of color with local conditions, the
nearly identical colors of binary KBOs argues that colors are simply primordial.
If binary KBOs were formed by early mutual capture in a 
quiescent disk \citep{2002Natur.420..643G}, 
the two component would likely have formed in very similar locations. If,
alternatively, binary KBOs were formed in an initial gravitational
collapse \citep{2010AJ....140..785N}, the objects would of necessity have formed at the 
same location and of the same materials. 

{\it Albedos.} The albedos of KBOs are an important indicator of surface
compositions. 
While measuring albedos of distant KBOs is difficult
and reliable results are known for mainly the largest 
(and thus perhaps most unrepresentative) objects, closer, warmer
centaurs have better characterized albedos. One result that 
appears robust is that the optically red centaurs have higher 
albedos than the optically blue centaurs, with the red centaur
albedos clustering around 4\% and the blue centaur albedos
a much higher 11\% \citep{2006ApJ...643..556S}.

\section{The volatile-loss gradient hypothesis}
If current Kuiper belt colors were set primarily by their formation 
location, as the similar colors of binary components and lack of correlation
of color with current environment suggests, the fact
that KBOs of all colors are now present throughout the Kuiper belt
strongly argues that substantial mixing has occurred after formation. 
Instability models such as the Nice model 
\citep{2005Natur.435..459T, 2008Icar..196..258L} 
provide natural 
mechanisms for this mixing, as objects which formed in the closer and more
massive primordial disk and dispersed essentially randomly throughout 
the regions that are now the Kuiper belt. 

A simplistic hypothesis, then, is that the expected chemical gradients
in the primordial disk \citep{2007prpl.conf..751B} lead to gradients
in the composition of the solid bodies. 
The large sold bodies, however, will move
differentially through the disk while growing, 
incorporating materials from diverse regions 
\citep{1997Icar..127..290W}, so the clear
chemical boundries of the disk are unlikely to be maintained
in the macroscopic bodies.
Indeed, short period comets,
derived from the Kuiper belt, show strong evidence for a chemical
diversity 
\citep[i.e.][]{2004come.book..391B, 2009ApJ...703..187D}.
which is unexplainable as a simple difference in formation location

In light of the large variability of cometary nuclei, we 
seek a hypothesis for the colors of the KBOs in which sharp color divisions
can occur even in the presence of substantial chemical variability. 
We consider the 
possibility of a surface evaporation gradient in the disk. In our scenario 
KBOs form out of a variable mix of cometary materials
and, as they are exposed to sunlight when the dust and gas disk first disappear,
their surfaces evolve. Depending on the heliocentric distance dependent thermal
equilibrium temperature, different molecules will evaporate and be lost
to space. Surfaces of 
objects further from the sun will retain more volatile 
molecules that are lost on the surfaces of objects closer to the sun.
When a Nice-like instability occurs and
the objects are scattered further from the sun, their temperatures
drop and evaporation of the less volatile species ceases.
Long term irradiation develops the colors currently seen on the surfaces
through chemical modification of whatever remains.

To quantitatively explore this scenario we use the volatile loss model
of \citet{2007ApJ...659L..61S} and examine evaporation as a function
of object size and heliocentric distance. We calculate the
minimum loss of molecular species by assuming Jeans escape -- the
slowest possible escape mechanism -- throughout the primordial disk. We
calculate the size-dependent heliocentric distance at which all of
a particular volatile should be depleted (Fig. 1). 

Because
of the exponential nature of Jeans escape, 
most specific model parameters chosen
have little impact on the final result, but for concreteness we
describe them here. We chose all parameters to be 
within the range of plausible values.
Objects are assumed to have  densities of 1.5 g cm$^{-3}$,
albedos are chosen to represent initially
frosty objects and are identical to Pluto, 
starting compositions are those measured in the 
Hale-Bopp coma \citep{2004come.book..391B}, with 
inert rock to give the correct final density. Even order-of-magnitude 
variations in the abundances do not change the 
conclusions below. We consider all
species with measured abundances relative to water of 0.5\% 
and
higher, with the exception of H$_2$CO, for which no accurate
laboratory data on the vapor pressure over the solid exists.
All other vapor pressure are taken from the compilation of
\citet{2009P&SS...57.2053F}.
The sun during this
early stage is assume to have 70\% of the current energy output
\citep{1981SoPh...74...21G}.

Two model parameters can affect the final result substantially.
First, 
we can either assume that
the entire object is porous and that volatiles throughout the body 
have access to the surface and are capable of eventual evaporation, or we
can assume that only a layer near the surface can evaporate. The final
colors of Kuiper belt objects depend only on the chemistry in a very
small layer near the surface, so both types of objects would appear identical
on the surface,
but the amount of material that needs to be evaporated for an object
to appear depleted differs greatly in these two scenarios. Even very
small short
period comets still appear to have
volatiles such as CO, which would quickly evaporate in the surface
layers of almost any object in the primordial Kuiper belt
\citep{2004come.book..391B}. We thus assume that evaporation is only
a surficial process and we set the depth required to deplete the
surface to 100 m.
Changing this depth by orders
of magnitude in either direction does not qualitatively change the conclusions
below.

The second important assumption is the length of time that passes from 
when the disk dissipates and the objects are exposed to sunlight
to when the objects are scattered to their more distant
locations. In the canonical
 Nice model, the scattering by the planetary instability
is the cause of the Late Heavy Bombardment 650 Myr after the formation
of the solar system \citep{2005Natur.435..466G}, 
so the primordial objects spend a long period
of time closer to the sun. Such a long period is not a requirement of
an instability model, however, so the actual time remains unconstrained.
Regardless of the exposure time, an irradiated crust can develop 
much more quickly \citep{2008ssbn.book..507H}, 
so additional exposure time can no longer affect colors.
We set our exposure time to be 10 Myr, but, again, values differing by an
order of magnitude in either direction do not qualitatively change the 
conclusions below.

For a wide range of assumptions, evaporation in the early Kuiper belt
behaves as shown in Figure 1. Water ice is involatile at all distances and
CO$_2$ and H$_2$S are involatile throughout the Kuiper belt.
Objects residing in the outer parts of the Kuiper belt retain
CH$_3$OH, then C$_2$H$_2$, C$_2$H$_6$, and HCN over a small range of
distances.
NH$_3$ is retained only near
the distance of the current cold classical Kuiper belt, and CH$_4$, N$_2$,
and CO are depleted on surface layers throughout the Kuiper belt except
for the largest objects.
For
the specific parameters chosen, the evaporation line of CH$_3$OH
appears near 20 AU, which would be in the
middle of the primordial disk of KBOs.
\begin{figure}
\plotone{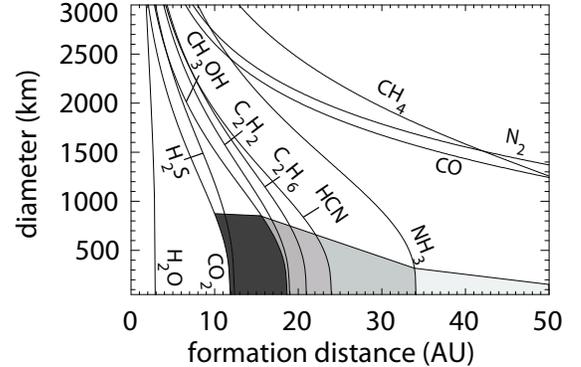}
\caption{Evaporation gradients on surfaces in the early solar system. Objects 
to the right of the lines have that molecular species depleted from
a 100 m layer in 10 Myr Changing the size of the depleted region or
the depletion time scale by an order of magnitude only moves the locations
of the curves by approximately an AU. 
The shaded regions show the formation regions of the three main surface
types of Kuiper belt objects. In the inner part of the Kuiper belt, 
only H$_2$O and CO$_2$ survive on the surface. Irradiation of these
molecules causes a dark neutral surface. In the middle part of 
the primordial Kuiper belt CH$_3$OH is retained on the surface, leading
to higher albedo redder surfaces. In the region of the current cold
classical Kuiper belt, NH$_3$ is retained, leading to the unique 
surface characteristics of these objects.
}
\end{figure}

\section{Surface Colors}
Experiments 
on ice irradiation in the outer solar system have primarily focused 
on specific chemical pathways and species rather than less
precise coloration \citep{2008ssbn.book..507H}, 
nonetheless some
trends appear clear.

Objects that form between the inner edge of the primordial Kuiper belt and
approximately 20 AU will have temperatures sufficiently high that the only
major ices that remain on the surface are H$_2$O, CO$_2$, and (with
a much smaller abundance) H$_2$S. No specific experiments have
been done on the coloration or albedo of such a mixture after
irradiation, but irradiation of H$_2$O, CO$_2$ mixtures is
known to produce carbonic acid and more complex hydrocarbons
\citep{1991JGR....9617541M, 1998JGR...10331391D}.
Irradiation of such hydrocarbons then leads to 
the loss of hydrogen, the production of larger carbon chains, and
the eventual carbonization of the surface. The final product 
is a dark neutrally colored spectrally bland surface 
\citep{1987A&A...184..333A, 2004Icar..170..214M,2004AdSpR..33...49P}. 
Irradiation of a H$_2$O, CO$_2$ mix has been speculated to be
a cause of the very dark crust of Callisto \citep{1998JGR...103.8603M}.

Such surfaces describe the dark neutrally colored centaurs
well. In addition, with the exception of some of the larger
water ice rich KBOs, the neutral KBOs have low
albedos similar to those of the centaurs \citep{2006ApJ...643..556S}.
We propose that the objects that are now
the neutral-colored KBOs were formed
in the inner part of the primordial disk and scattered into the 
current Kuiper belt. When they scatter inward to become centaurs
their temperatures do not increase markedly beyond those that
were experiences at formation, so surfaces do not evolve significantly.

Beyond approximately 20 AU, several major hydrocarbon species can
remain on the surface. The most abundant of these is methanol. 
A Raman study of residues remaining after methanol irradiation
showed that the methanol residue is surprisingly lacking
in the signature of amorphous carbon \citep{2004A&A...414..757F}. 
Indeed, \citet{2006ApJ...644..646B}
have shown that methanol, when irradiated to dosages expected for
solar system aged KBOs \citep{2003EM&P...92..261C},
 does not turn dark and neutral  
but instead 
retains higher albedos and redder colors. The colors and albedos
are similar to those seen in the red centaurs and likely also
the medium-size red KBOS. Methanol, intriguingly, is also
the only involatile molecule other than water identified either on a 
centaur \citep{1998Icar..135..389C} or a KBO \citep{2006A&A...455..725B}. 
We propose that the presence of
methanol on the primordial surface of a KBO allows that KBO to 
maintain a higher albedo redder irradiation crust. It is possible -- indeed
likely -- that that the C$_2$H$_6$ and the C$_2$H$_2$ and HCN evaporation lines,
which
are just beyond
 the methanol line effect colors also. We thus expect the red objects
to exhibit a wider range of colors than the more uniform neutral objects.
These red objects, when scattered into the centaur region, also need not
evolve significantly, as the red irradiation mantle is stable to
higher temperature than pure methanol.

The neutral and red KBOs scatter throughout the outer solar system, and in 
the classical Nice model they are responsible for populating
 the cold classical Kuiper belt
\citep{2008Icar..196..258L}. 
Such a formation is inconsistent with the observation that the cold classical
objects are composed purely of red objects and 
have other unique characteristics.
Recently, however, 
Batygin et al. (2011) have shown that the cold classical belt
need not have been emplaced during the Nice instability but rather 
could have survived if formed in place.
Objects formed beyond 35 AU would have been at all times 
beyond the ammonia evaporation line;
No laboratory experiments
have tested whether the addition of 
ammonia will have a significant effect on coloration after irradiation,
but ammonia has been shown to block some types of irradiation 
chemistry \citep{2001JGR...10633275H}, which could possibly
affect coloration.
Ammonia is the only major species to have an evaporation line
outside of the expected primordial Kuiper belt but inside of the 
current cold classical Kuiper belt. We thus propose that 
the addition of ammonia is the cause of the unique colors and
high albedos of the cold classical KBOs.

In our scenario, little evolution occurs as KBOs move into the centaur
region. The non-cold classical KBOs should, then, have 
an identical color distribution as the centaurs.
We examine whether the updated
ground based color database\footnote{Compiled at http://www.eso.org/$\sim$ohainaut/MBOSS} of
\citet{2002A&A...389..641H}
supports this implication. 
We select all objects with
high quality color measurements (color gradient errors under 10\%) and 
absolute magnitudes between $H=6$ and $H=9$ (where measurements
exist of both KBOs and centaurs, so size-related effects are minimized).
 We further separate those into the 19
with perihelion inside of 20 AU, to represent centaurs which
could have thermally evolved and all 83 non-cold classical objects with
perihelion greater than 30 AU, to represent an unevolved
population. Histograms of these color distributions are show in Figure 2. 
The distributions do not appear to be significantly different. Indeed,
a K-S test shows that the two populations are
distinguishable only at the 48\% confidence level, or, to state the result
clearly, there is no evidence
from the ground-based data that the centaur and the non-cold classical
KBO colors are drawn from different populations. While this result sounds
surprising given the long history of comparison of colors of centaurs
and KBOs \citep[reviewed in][]{2008ssbn.book...91D}, this is the first test
of this specific subpopulation. The fact that these two populations
are currently indistinguishable does not mean, of course, that they
are identical. Further color observations will be crucial to 
further test this prediction.
\begin{figure}
\plotone{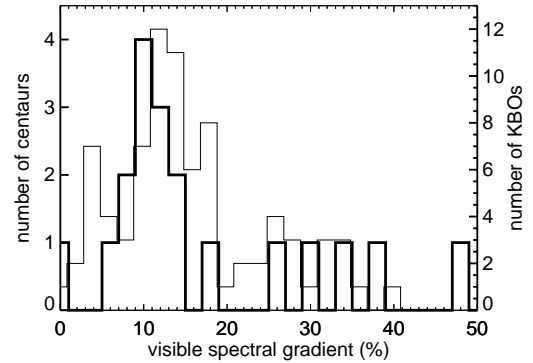}
\caption{A comparison between the visible colors of centaurs (thick lines) and
of similarly-sized non-cold classical KBOs (thin lines). A K-S test cannot
distinguish any difference between these populations. While this
statistical lack of difference is intriguing, more color data 
are needed --
particularly of centaurs in this size range -- in order to determine if
the distributions are truly identical.}
\end{figure}

\section{Discussion}
We propose that evaporation of volatiles and subsequent irradiation
of the remaining ices 
are the basis for the diverse colors of the Kuiper belt. The existence
of three major zones of volatile evaporation in the primordial Kuiper
belt gives rise to three major types of surface compositions. Laboratory
experiments, when available, support the correlation of the colors and 
albedo with the specific volatiles expected to be retained on the surface.

The hypothesis advanced here is necessarily non-unique. While the key
physical process of evaporation of volatiles is inevitable, the interaction
between the surface, the subsurface, the atmosphere, and space is
sufficiently complex and unknown that more detailed modeling would
give little additional insight. The few direct observational
tests of this model test, at best, aspects only peripherally
related, nonetheless, we predict:
(1) Precision photometric observations will continue to show that the
non-cold classical KBOs form two broad color groupings consistent
with the centaurs of similar size;
(2) With the exception of large KBOs and Haumea family members, all
neutrally colored KBOs are similar in surface ice composition
and will have the low albedos of the neutral centaurs; high precision 
albedo measurements may additionally show the small expected
systematic albedo variation 
due to mixing with a non-ice component;
(3) red KBOs
will show slightly greater variability, but, with the exception
of the largest objects, they will also not show a large range of
albedos, but will have the elevated albedos of the red centaurs.

In lieu of powerful observational or theoretical
tests, we expect that laboratory experiments will be the
most useful for supporting or refuting this hypothesis. The
studies used to suggest the chemical and color behavior here 
have been performed in a wide variety of laboratories concentrating
on diverse subjects. For the hypothesis to be supported, a systematic
laboratory study would have to show the following effects:
(1)  Water - carbon dioxide (and, possible H$_2$S) ice mixtures 
must carbonize to form a
neutrally-colored low albedo surface when irradiated at the appropriate
level; (2) the addition of methanol to the ice mix must allow the
irradiation product to remain red and have a higher albedo; 
and (3) the addition of ammonia to the mix
must reproduce the unique colors and albedos
of the cold classical Kuiper belt objects.
While all of these possibilities are supported by the current laboratory
data, only experimental verification will allow support or refutation
of this hypothesis for explaining the colors of objects in the Kuiper belt.

\acknowledgements This research has been supported by grant
NNX09AB49G from the NASA Planetary Astronomy program. 
We thank Hal Levison, Rosario Brunetto, and Konstantin Batygin
 for enlightening
conversations.


\begin{thebibliography}{41}
\expandafter\ifx\csname natexlab\endcsname\relax\def\natexlab#1{#1}\fi

\bibitem[{{Morbidelli} \& {Brown}(2005)}]{morbandbrown}
{ Morbidelli}, A. \& {Brown}, M.~E. 2005, {Comets II, ed. M.C. Festou, H.U.
  Keller, H.A. Weaver (U. Arizona Press: 2005)}

\bibitem[{{Andronico} {et~al.}(1987){Andronico}, {Baratta}, {Spinella}, \&
  {Strazzulla}}]{1987A&A...184..333A}
{Andronico}, G., {Baratta}, G.~A., {Spinella}, F., \& {Strazzulla}, G. 1987,
  \aap, 184, 333

\bibitem[{{Barucci} {et~al.}(2006){Barucci}, {Merlin}, {Dotto},
  {Doressoundiram}, \& {de Bergh}}]{2006A&A...455..725B}
{Barucci}, M.~A., {Merlin}, F., {Dotto}, E., {Doressoundiram}, A., \& {de
  Bergh}, C. 2006, \aap, 455, 725

\noindent Batygin, K., Brown, M.E., \& Fraser, W.C., 2011, \apj, in press

\bibitem[{{Benecchi} {et~al.}(2009){Benecchi}, {Noll}, {Grundy}, {Buie},
  {Stephens}, \& {Levison}}]{2009Icar..200..292B}
{Benecchi}, S.~D., {Noll}, K.~S., {Grundy}, W.~M., {Buie}, M.~W., {Stephens},
  D.~C., \& {Levison}, H.~F. 2009, Icarus, 200, 292

\bibitem[{{Bergin} {et~al.}(2007){Bergin}, {Aikawa}, {Blake}, \& {van
  Dishoeck}}]{2007prpl.conf..751B}
{Bergin}, E.~A., {Aikawa}, Y., {Blake}, G.~A., \& {van Dishoeck}, E.~F. 2007,
  Protostars and Planets V, 751

\bibitem[{{Bockel{\'e}e-Morvan} {et~al.}(2004){Bockel{\'e}e-Morvan},
  {Crovisier}, {Mumma}, \& {Weaver}}]{2004come.book..391B}
{Bockel{\'e}e-Morvan}, D., {Crovisier}, J., {Mumma}, M.~J., \& {Weaver}, H.~A.
  {The composition of cometary volatiles}, ed. {Festou, M.~C., Keller, H.~U.,
  \& Weaver, H.~A.}, 391--423

\bibitem[{{Brown}(2001)}]{2001AJ....121.2804B}
{Brown}, M.~E. 2001, \aj, 121, 2804

\bibitem[{{Brown} {et~al.}(2007){Brown}, {Barkume}, {Ragozzine}, \&
  {Schaller}}]{2007Natur.446..294B}
{Brown}, M.~E., {Barkume}, K.~M., {Ragozzine}, D., \& {Schaller}, E.~L. 2007,
  \nat, 446, 294

\bibitem[{{Brucker} {et~al.}(2009){Brucker}, {Grundy}, {Stansberry}, {Spencer},
  {Sheppard}, {Chiang}, \& {Buie}}]{2009Icar..201..284B}
{Brucker}, M.~J., {Grundy}, W.~M., {Stansberry}, J.~A., {Spencer}, J.~R.,
  {Sheppard}, S.~S., {Chiang}, E.~I., \& {Buie}, M.~W. 2009, Icarus, 201, 284

\bibitem[{{Brunetto} {et~al.}(2006){Brunetto}, {Barucci}, {Dotto}, \&
  {Strazzulla}}]{2006ApJ...644..646B}
{Brunetto}, R., {Barucci}, M.~A., {Dotto}, E., \& {Strazzulla}, G. 2006, \apj,
  644, 646

\bibitem[{{Cooper} {et~al.}(2003){Cooper}, {Christian}, {Richardson}, \&
  {Wang}}]{2003EM&P...92..261C}
{Cooper}, J.~F., {Christian}, E.~R., {Richardson}, J.~D., \& {Wang}, C. 2003,
  Earth Moon and Planets, 92, 261

\bibitem[{{Cruikshank} {et~al.}(1998){Cruikshank}, {Roush}, {Bartholomew},
  {Geballe}, {Pendleton}, {White}, {Bell}, {Davies}, {Owen}, {de Bergh},
  {Tholen}, {Bernstein}, {Brown}, {Tryka}, \& {Dalle
  Ore}}]{1998Icar..135..389C}
{Cruikshank}, D.~P., {Roush}, T.~L., {Bartholomew}, M.~J., {Geballe}, T.~R.,
  {Pendleton}, Y.~J., {White}, S.~M., {Bell}, J.~F., {Davies}, J.~K., {Owen},
  T.~C., {de Bergh}, C., {Tholen}, D.~J., {Bernstein}, M.~P., {Brown}, R.~H.,
  {Tryka}, K.~A., \& {Dalle Ore}, C.~M. 1998, \icarus, 135, 389

\bibitem[{{Delitsky} \& {Lane}(1998)}]{1998JGR...10331391D}
{Delitsky}, M.~L. \& {Lane}, A.~L. 1998, \jgr, 103, 31391

\bibitem[{{Dello Russo} {et~al.}(2009){Dello Russo}, {Vervack}, {Weaver},
  {Kawakita}, {Kobayashi}, {Biver}, {Bockel{\'e}e-Morvan}, \&
  {Crovisier}}]{2009ApJ...703..187D}
{Dello Russo}, N., {Vervack}, Jr., R.~J., {Weaver}, H.~A., {Kawakita}, H.,
  {Kobayashi}, H., {Biver}, N., {Bockel{\'e}e-Morvan}, D., \& {Crovisier}, J.
  2009, \apj, 703, 18

\bibitem[{{Doressoundiram} {et~al.}(2008){Doressoundiram}, {Boehnhardt},
  {Tegler}, \& {Trujillo}}]{2008ssbn.book...91D}
{Doressoundiram}, A., {Boehnhardt}, H., {Tegler}, S.~C., \& {Trujillo}, C.
  {Color Properties and Trends of the Transneptunian Objects}, ed. {Barucci,
  M.~A., Boehnhardt, H., Cruikshank, D.~P., Morbidelli, A., \& Dotson, R.},
  91--104

\bibitem[{{Ferini} {et~al.}(2004){Ferini}, {Baratta}, \&
  {Palumbo}}]{2004A&A...414..757F}
{Ferini}, G., {Baratta}, G.~A., \& {Palumbo}, M.~E. 2004, \aap, 414, 757

\bibitem[{{Fraser} {et~al.}(2010){Fraser}, {Brown}, \&
  {Schwamb}}]{2010Icar..210..944F}
{Fraser}, W.~C., {Brown}, M.~E., \& {Schwamb}, M.~E. 2010, \icarus, 210, 944

\bibitem[{{Fray} \& {Schmitt}(2009)}]{2009P&SS...57.2053F}
{Fray}, N. \& {Schmitt}, B. 2009, \planss, 57, 2053

\bibitem[{{Goldreich} {et~al.}(2002){Goldreich}, {Lithwick}, \&
  {Sari}}]{2002Natur.420..643G}
{Goldreich}, P., {Lithwick}, Y., \& {Sari}, R. 2002, \nat, 420, 643

\bibitem[{{Gomes} {et~al.}(2005){Gomes}, {Levison}, {Tsiganis}, \&
  {Morbidelli}}]{2005Natur.435..466G}
{Gomes}, R., {Levison}, H.~F., {Tsiganis}, K., \& {Morbidelli}, A. 2005, \nat,
  435, 466

\bibitem[{{Gough}(1981)}]{1981SoPh...74...21G}
{Gough}, D.~O. 1981, \solphys, 74, 21

\bibitem[{{Hainaut} \& {Delsanti}(2002)}]{2002A&A...389..641H}
{Hainaut}, O.~R. \& {Delsanti}, A.~C. 2002, \aap, 389, 641

\bibitem[{{Hudson} \& {Moore}(2001)}]{2001JGR...10633275H}
{Hudson}, R.~L. \& {Moore}, M.~H. 2001, \jgr, 106, 33275

\bibitem[{{Hudson} {et~al.}(2008){Hudson}, {Palumbo}, {Strazzulla}, {Moore},
  {Cooper}, \& {Sturner}}]{2008ssbn.book..507H}
{Hudson}, R.~L., {Palumbo}, M.~E., {Strazzulla}, G., {Moore}, M.~H., {Cooper},
  J.~F., \& {Sturner}, S.~J. {Laboratory Studies of the Chemistry of
  Transneptunian Object Surface Materials}, ed. {Barucci, M.~A., Boehnhardt,
  H., Cruikshank, D.~P., Morbidelli, A., \& Dotson, R.}, 507--523

\bibitem[{{Jewitt} \& {Luu}(1998)}]{1998AJ....115.1667J}
{Jewitt}, D. \& {Luu}, J. 1998, \aj, 115, 1667

\bibitem[{{Levison} {et~al.}(2008){Levison}, {Morbidelli}, {Vanlaerhoven},
  {Gomes}, \& {Tsiganis}}]{2008Icar..196..258L}
{Levison}, H.~F., {Morbidelli}, A., {Vanlaerhoven}, C., {Gomes}, R., \&
  {Tsiganis}, K. 2008, \icarus, 196, 258

\bibitem[{{Levison} \& {Stern}(2001)}]{2001AJ....121.1730L}
{Levison}, H.~F. \& {Stern}, S.~A. 2001, \aj, 121, 1730

\bibitem[{{Luu} \& {Jewitt}(1996)}]{1996AJ....112.2310L}
{Luu}, J. \& {Jewitt}, D. 1996, \aj, 112, 2310

\bibitem[{{McCord} {et~al.}(1998){McCord}, {Hansen}, {Clark}, {Martin},
  {Hibbitts}, {Fanale}, {Granahan}, {Segura}, {Matson}, {Johnson}, {Carlson},
  {Smythe}, \& {Danielson}}]{1998JGR...103.8603M}
{McCord}, T.~B., {Hansen}, G.~B., {Clark}, R.~N., {Martin}, P.~D., {Hibbitts},
  C.~A., {Fanale}, F.~P., {Granahan}, J.~C., {Segura}, M., {Matson}, D.~L.,
  {Johnson}, T.~V., {Carlson}, R.~W., {Smythe}, W.~D., \& {Danielson}, G.~E.
  1998, \jgr, 103, 8603

\bibitem[{{Moore} {et~al.}(1991){Moore}, {Khanna}, \&
  {Donn}}]{1991JGR....9617541M}
{Moore}, M.~H., {Khanna}, R., \& {Donn}, B. 1991, \jgr, 96, 17541

\bibitem[{{Moroz} {et~al.}(2004){Moroz}, {Baratta}, {Strazzulla}, {Starukhina},
  {Dotto}, {Barucci}, {Arnold}, \& {Distefano}}]{2004Icar..170..214M}
{Moroz}, L., {Baratta}, G., {Strazzulla}, G., {Starukhina}, L., {Dotto}, E.,
  {Barucci}, M.~A., {Arnold}, G., \& {Distefano}, E. 2004, \icarus, 170, 214

\bibitem[{{Nesvorn{\'y}} {et~al.}(2010){Nesvorn{\'y}}, {Youdin}, \&
  {Richardson}}]{2010AJ....140..785N}
{Nesvorn{\'y}}, D., {Youdin}, A.~N., \& {Richardson}, D.~C. 2010, \aj, 140, 785

\bibitem[{{Noll} {et~al.}(2008){Noll}, {Grundy}, {Stephens}, {Levison}, \&
  {Kern}}]{2008Icar..194..758N}
{Noll}, K.~S., {Grundy}, W.~M., {Stephens}, D.~C., {Levison}, H.~F., \& {Kern},
  S.~D. 2008, Icarus, 194, 758

\bibitem[{{Palumbo} {et~al.}(2004){Palumbo}, {Ferini}, \&
  {Baratta}}]{2004AdSpR..33...49P}
{Palumbo}, M.~E., {Ferini}, G., \& {Baratta}, G.~A. 2004, Advances in Space
  Research, 33, 49

\bibitem[{{Schaller} \& {Brown}(2007)}]{2007ApJ...659L..61S}
{Schaller}, E.~L. \& {Brown}, M.~E. 2007, \apjl, 659, L61

\bibitem[{{Schaller} {et~al.}(2009){Schaller}, {Brown}, \&
  {Haghighipour}}]{2009AGUFM.P21B1213S}
{Schaller}, E.~L., {Brown}, M.~E., \& {Haghighipour}, N. 2009, AGU Fall Meeting
  Abstracts, B1213+

\bibitem[{{Stansberry} {et~al.}(2006){Stansberry}, {Grundy}, {Margot},
  {Cruikshank}, {Emery}, {Rieke}, \& {Trilling}}]{2006ApJ...643..556S}
{Stansberry}, J.~A., {Grundy}, W.~M., {Margot}, J.~L., {Cruikshank}, D.~P.,
  {Emery}, J.~P., {Rieke}, G.~H., \& {Trilling}, D.~E. 2006, \apj, 643, 556

\bibitem[{{Stern}(2002)}]{2002AJ....124.2297S}
{Stern}, S.~A. 2002, \aj, 124, 2297

\bibitem[{{Tegler} {et~al.}(2008){Tegler}, {Bauer}, {Romanishin}, \&
  {Peixinho}}]{2008ssbn.book..105T}
{Tegler}, S.~C., {Bauer}, J.~M., {Romanishin}, W., \& {Peixinho}, N. {Colors of
  Centaurs}, ed. {Barucci, M.~A., Boehnhardt, H., Cruikshank, D.~P.,
  Morbidelli, A., \& Dotson, R.}, 105--114

\bibitem[{{Trujillo} \& {Brown}(2002)}]{2002ApJ...566L.125T}
{Trujillo}, C.~A. \& {Brown}, M.~E. 2002, \apjl, 566, L125

\bibitem[{{Tsiganis} {et~al.}(2005){Tsiganis}, {Gomes}, {Morbidelli}, \&
  {Levison}}]{2005Natur.435..459T}
{Tsiganis}, K., {Gomes}, R., {Morbidelli}, A., \& {Levison}, H.~F. 2005, \nat,
  435, 459

\bibitem[{{Weidenschilling}(1997)}]{1997Icar..127..290W}
{Weidenschilling}, S.~J. 1997, \icarus, 127, 290

\end{thebibliography}
\end{document}